\documentclass{ws-p10x7}
\begin{document}
\newcommand {\ignore}[1]{}
\def\tp{these proceedings}
\def\smallfrac#1#2{{\textstyle{#1 \over #2}}}
\def\lsim{\:\raisebox{-0.5ex}{$\stackrel{\textstyle<}{\sim}$}\:}
\def\gsim{\:\raisebox{-0.5ex}{$\stackrel{\textstyle>}{\sim}$}\:}
\def\VEV#1{\left\langle #1\right\rangle}
\def\fig#1{{Fig. (\ref{#1})}}
\def\smallfrac#1#2{{\textstyle{#1 \over #2}}}
\def\thefootnote{\fnsymbol{footnote}}
\def\N{$\cal N$ }
\def\slash#1{#1\!\!\! /}
\def\nn{\nonumber}
\def\rp{$R_p \hspace{-1em}/\;\:$ }
\def\Eq#1{{Eq. (\ref{#1})}}
\def\eq#1{{eq. (\ref{#1})}}
\def\Fig#1{{Fig. (\ref{#1})}}
\def\be{\begin{equation}}
\def\ee{\end{equation}}    
\def\bear{\be\begin{array}}
\def\eear{\end{array}\ee}
\def\bea{\begin{eqnarray}}
\def\eea{\end{eqnarray}}
\def\SM{Standard Model }
\def\baselinestretch{1.2}
\def\vb#1{\vbox to #1 pt{}}
\def\beqa{\begin{eqnarray}}
\def\eeqa{\end{eqnarray}}
\def\ni{\noindent}
\def\ba{\begin{array}}
\def\ea{\end{array}}
\def\ovl{\overline}
\def\ds{\displaystyle}
\def\epjc#1#2#3{{\it Eur.\ Phys.\ J. }{\bf C #1} (#2) #3}
\def\npb#1#2#3{{\it Nucl.\ Phys.\ }{\bf B #1} (#2) #3}
\def\plb#1#2#3{{\it Phys.\ Lett.\ }{\bf B #1} (#2) #3}  
\def\prd#1#2#3{{\it Phys.\ Rev.\ }{\bf D #1} (#2) #3}
\def\prep#1#2#3{{\it Phys.\ Rep.\ }{\bf #1} (#2) #3}
\def\prl#1#2#3{{\it Phys.\ Rev.\ Lett.\ }{\bf #1} (#2) #3}
\def\mpla#1#2#3{{\it Mod.\ Phys.\ Lett.\ }{\bf A #1} (#2) #3}
\def\sjnp#1#2#3{{\it Sov.\ J.\ Nucl.\ Phys.\ }{\bf #1} (#2) #3}
\def\jetpl#1#2#3{{\it Sov.\ Phys.\ JETP Lett.\ }{\bf #1} (#2) #3}
\def\rnc#1#2#3{{\it Riv. Nuovo Cimento }{\bf #1} (#2) #3}
\def\yf#1#2#3{{\it Yad.\ Fiz.\ }{\bf #1} (#2) #3}
\def\hepph#1{{\tt hep-ph/#1}}
\def\ne{\hbox{$\nu_e$ }}
\def\nm{\hbox{$\nu_\mu$ }}
\def\nt{\hbox{$\nu_\tau$ }}        
\def\21{$SU(2) \otimes U(1)$}
\def\ie{{\it i.e.}}
\def\etal{{\it et al.}}
\def\half{{\textstyle{1 \over 2}}}
\def\third{{\textstyle{1 \over 3}}}
\def\quarter{{\textstyle{1 \over 4}}}
\def\sixth{{\textstyle{1 \over 6}}}
\def\eighth{{\textstyle{1 \over 8}}}
\def\sqrthalf{{\textstyle{1 \over \sqrt{2}}}}
\def\bold#1{\setbox0=\hbox{$#1$}
     \kern-.025em\copy0\kern-\wd0
     \kern.05em\copy0\kern-\wd0
     \kern-.025em\raise.0433em\box0 }

 \newcommand{\wt}{\widetilde}
 \newcommand {\chiz} [1] {\tilde{\chi}^{0}_{#1} }
 \newcommand {\chiw} [1] {\tilde{\chi}^{\pm}_{#1} }

\def\rp{$R_p \hspace{-1em}/\;\:$ }
\def\Eq#1{{Eq. (\ref{#1})}}
\def\eq#1{{eq. (\ref{#1})}}
\def\beqa{\begin{eqnarray}}
\def\eeqa{\end{eqnarray}}

\def\vb#1{\vbox to #1 pt{}}

\title{ Broken R Parity, Neutrino Anomalies and Collider Tests}

\author{M. Hirsch, W. Porod, J. Rom\~ao $^*$ \& \underline{J.~W.~F.~Valle}}

\address{Instituto de F\'{\i}sica Corpuscular -- C.S.I.C. -- Universitat de
 Val\`encia  \\
     Ed. de Institutos de Paterna -- Apartado de Correos 22085 -
    46071  Val\`encia, Spain}
  \address{$^*$ Inst. Superior Tecnico, Depto. de Fisica, Av.
    Rovisco Pais, 1 1096 Lisboa Codex, Portugal}

\twocolumn[\maketitle\abstract{
  
  The solar and atmospheric neutrino anomalies constitute the only
  solid and most remarkable evidence for physics beyond the Standard
  Model, indicating that the lepton mixing matrix is fundamentally
  distinct from that describing the quarks.  Here I will report on how
  supersymmetry with spontaneously or bilinearly broken R Parity
  provides a predictive theory for neutrino mass and mixing which
  leads to a solution of neutrino anomalies which can be clearly
  tested at high energy accelerators.}]

\section{Motivation}

The simplest interpretation of the solar and atmospheric neutrino
data~\cite{update00,Fornengo:2000sr,MSW00} indicate that, in contrast
to quark mixing, possibly two of the lepton mixing angles are large.
Here I discuss how supersymmetry with broken R Parity provides
a predictive theoretical model for neutrino mass and mixing which
solves the solar and atmospheric neutrino anomalies in a way that
allows the leptonic mixing angles to be probed at high energy
accelerators.

R-parity conservation is an \texttt{ad hoc} assumption in the MSSM and
\rp may arise \texttt{explicitly} as unification remnant or
\texttt{spontaneously} by \21 doublet left sneutrino vacuum
expectation values (VEVS) $\VEV{\wt \nu_i}$ as originally
suggested~\cite{old,old2} but with an \texttt{ad hoc} set of explicit
breaking terms~\cite{Ross:1985yg} to comply with LEP data on Z width.
Preferably we break R-parity spontaneously through \texttt{singlet
  right sneutrino VEVS}, either by gauging L-number, in which case
there is an additional Z~\cite{Gonzalez-Garcia:1991qf} or within the
\21 scheme, in which case the \texttt{majoron} is an \21 singlet, with
suppressed Z coupling~\cite{SBRpV}. Spontaneous R-parity violation may
lead to a successful electroweak baryogenesis~\cite{Multamaki:1998fu}.

If R-parity is broken spontaneously then \texttt{only bilinear \rp
  terms arise in the effective theory} below the \rp violation scale.
Bilinear R--parity violation may also be assumed \texttt{ab initio} as
the fundamental theory. For example, it may be the only violation
permitted by higher Abelian flavour symmetries~\cite{Mira:2000gg}.
Moreover the bilinear model provides a theoretically self-consistent
scheme in the sense that trilinear \rp implies, by renormalization
group effects, that also bilinear \rp is present, but \texttt{not}
conversely.  The simplest \rp model (we call it \rp MSSM) is
characterized by three independent parameters in addition to those
specifying the minimal MSSM model. As shown in
ref.~\cite{Hirsch:2000ef} this leads to a predictive pattern for
neutrino masses and mixing angles which provides a solution to the
solar and atmospheric neutrino problems. It also predicts a well
specified pattern of \rp phenomena that can be tested at collider
experiments, providing an independent determination of neutrino mixing
angles at high energy accelerator experiments.

\section{ Bilinear \rp MSSM }
 
\begin{figure}
\setlength{\unitlength}{1mm}
\begin{picture}(60,33)
\put(0,-21){\mbox{\epsfig{figure=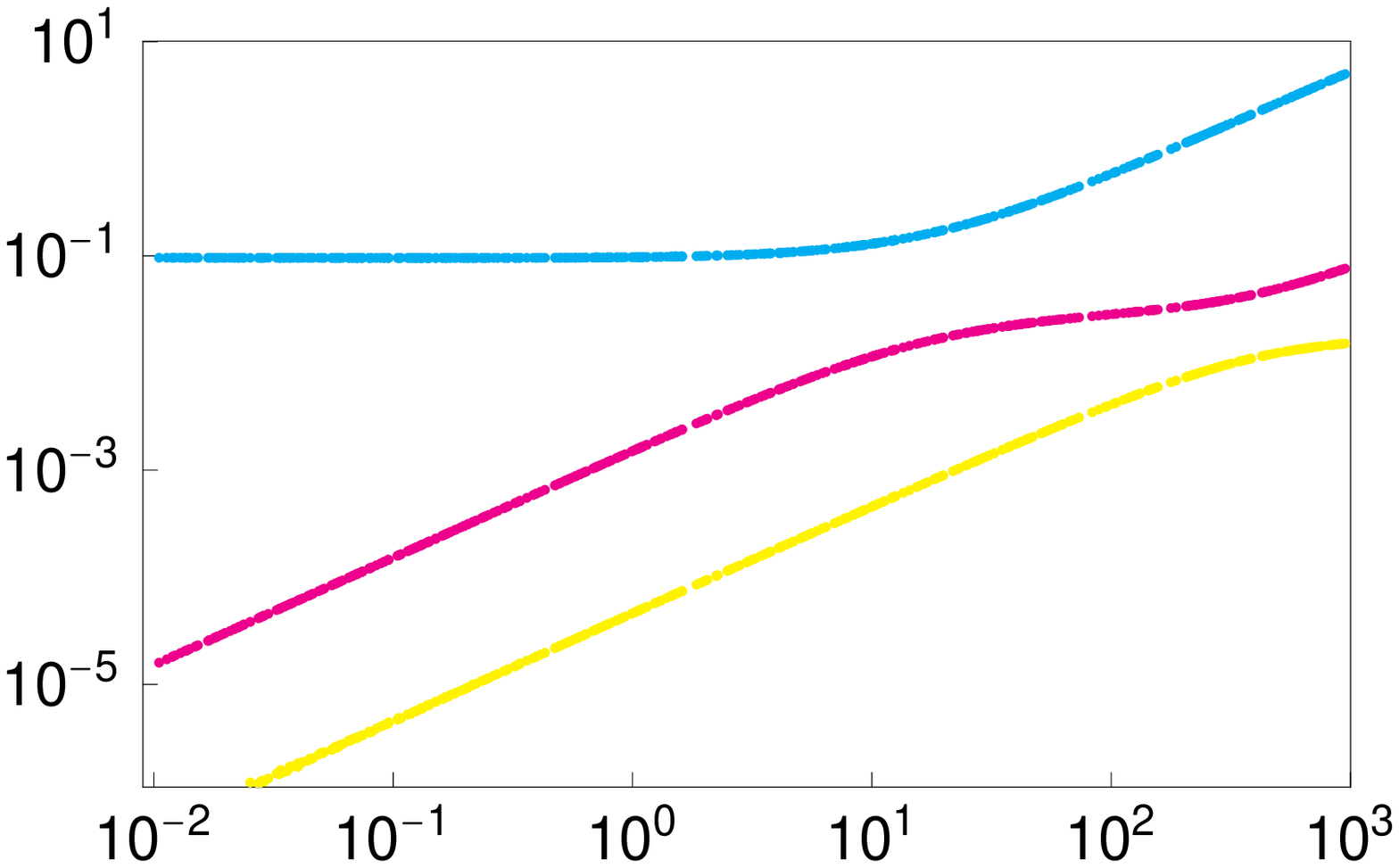,height=8cm,width=6cm}}}
\put(1,38){\mbox{{\small $m_{1,2,3}$~[eV]}}}
\put(48,-1){\mbox{{\small $|\epsilon|^2/|\Lambda|$}}}
\end{picture}
\caption{Typical neutrino masses in  \rp MSSM. }
\label{fig:masses}
\end{figure}

\begin{figure*}
\setlength{\unitlength}{1mm}
\begin{picture}(60,44)
\put(0,0){\mbox{\epsfig{figure=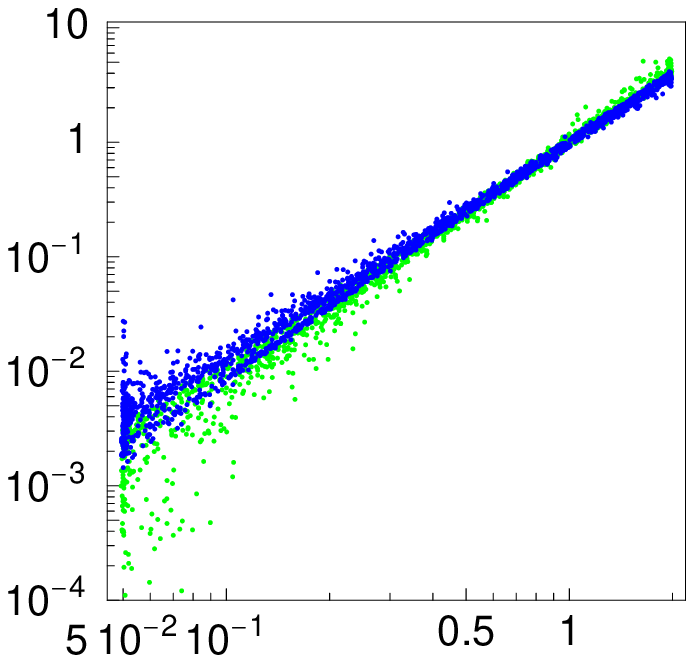,height=4.5cm,width=4.3cm}}}
\put(50,0){\mbox{\epsfig{figure=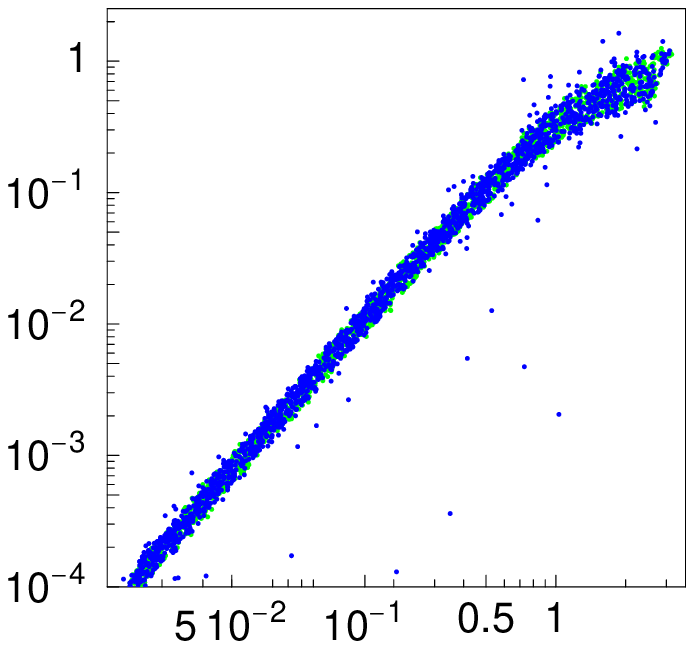,height=4.5cm,width=4.3cm}}}
\put(99,0){\mbox{\epsfig{figure=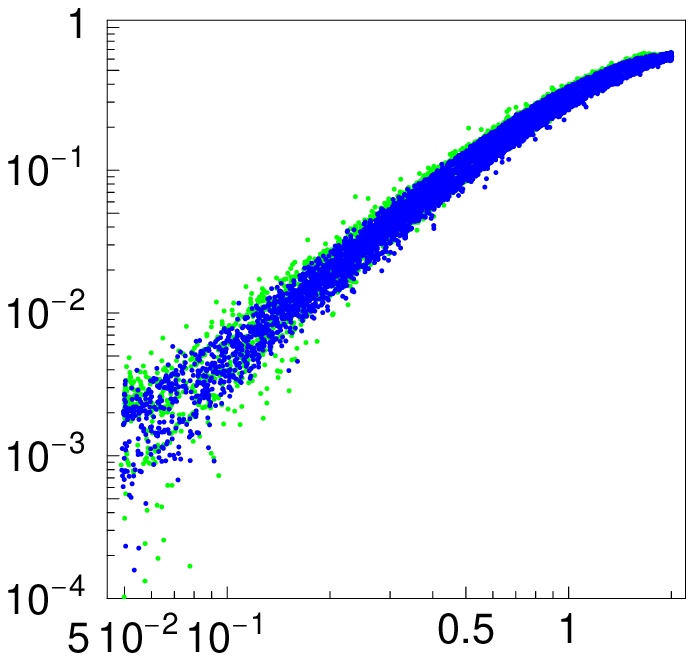,height=4.5cm,width=4.3cm}}}
\put(1,46){\mbox{$\tan^2 \theta_{atm}$}}
\put(15,-2){\mbox{$|\Lambda_\mu|/\sqrt{\Lambda^2_e+\Lambda^2_\tau}$}}
\put(52,46){\mbox{$\tan^2 \theta_{sol}$}}
\put(86,0){\mbox{$\epsilon_e/\epsilon_\mu$}}
\put(103,46){\mbox{$U^2_{e3}$}}
\put(115,-3){\mbox{$|\Lambda_e|/\sqrt{\Lambda^2_\mu+\Lambda^2_\tau}$}}
\end{picture}
\caption{Neutrino mixing angles in  \rp MSSM. \label{fig:angles}}
\end{figure*}

The minimal supergravity version of R-parity breaking
MSSM~\cite{Diaz:1998xc} is specified by the superpotential,
\begin{equation}  
W= W_{MSSM} + \epsilon_i L_i \widehat H_u 
\label{eq:Wsuppot} 
\end{equation} 
Since lepton number is broken, neutrinos pick up a mass. The expected
neutrino mass pattern is illustrated in \fig{fig:masses}, taken
from~\cite{Hirsch:2000ef}. It is typically hierarchical since only one
neutrino acquires mass at the tree level, while the others get mass
from calculable radiative corrections~\cite{Hirsch:2000ef}. As a
result neutrino masses can account for the solar and atmospheric
neutrino problems ~\footnote{For such small masses neutrinoless double
  beta decay has been shown to be too small to   observe~\cite{Hirsch:2000jt}}.
Having only bilinear R-parity violating terms as the origin of the
neutrino masses implies also that the three neutrino mixing angles
(assuming CP conservation in the lepton sector) are determined as
functions of the three bilinear \rp terms, leading to a predictive
scenario, independently of any particular form for the charged lepton
mass matrix. This is illustrated in \fig{fig:angles}, taken
from~\cite{Hirsch:2000ef}.
As can be seen, large angle solar solutions, LMA and LOW, now
preferred by solar spectrum data and by the global fit of all solar
neutrino data, as well as small angle solution (preferred by the
rates) can be accounted for within the theory. However, as explained
in ~\cite{Hirsch:2000ef}, for the very particular case of strictly
universal boundary conditions at the unification scale, consistency
with the reactor experiments~\cite{chooz} implies the SMA solar
solution.

\section{   Implications}

\begin{figure}
\setlength{\unitlength}{1mm}
\begin{picture}(60,35)
\put(0,-1){\mbox{\epsfig{figure=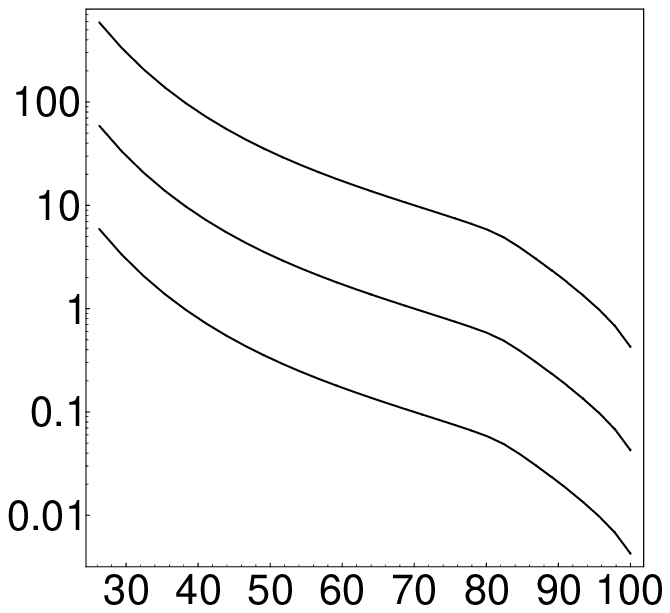,height=4cm,width=6.cm}}}
\put(3,39){\mbox{{\small L$(\chi^0_1)$~[cm]}}}
\put(33,28){\mbox{{\small 0.01}}}
\put(33,22){\mbox{{\small 0.1}}}
\put(33,16){\mbox{{\small 1}}}
\put(46,-1){\mbox{{\small $m_{\chi^0_1}$~[GeV]}}}
\end{picture}
\caption{Typical neutralino decay length in  \rp MSSM. }
\label{fig:dchi0}
\end{figure}

\begin{figure*}
\setlength{\unitlength}{1mm}
\begin{picture}(60,44)
\put(0,0){\mbox{\epsfig{
                   figure=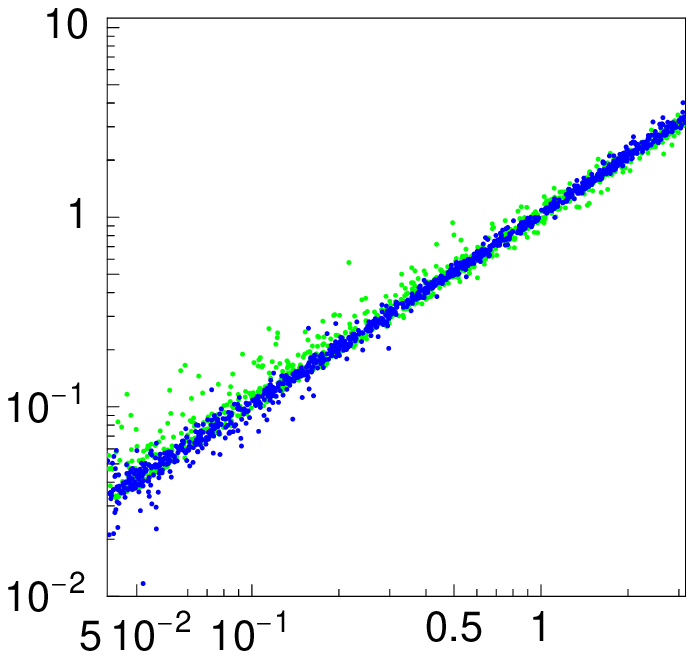,height=4.5cm,width=4.3cm}}}
\put(50,0){\mbox{\epsfig{figure=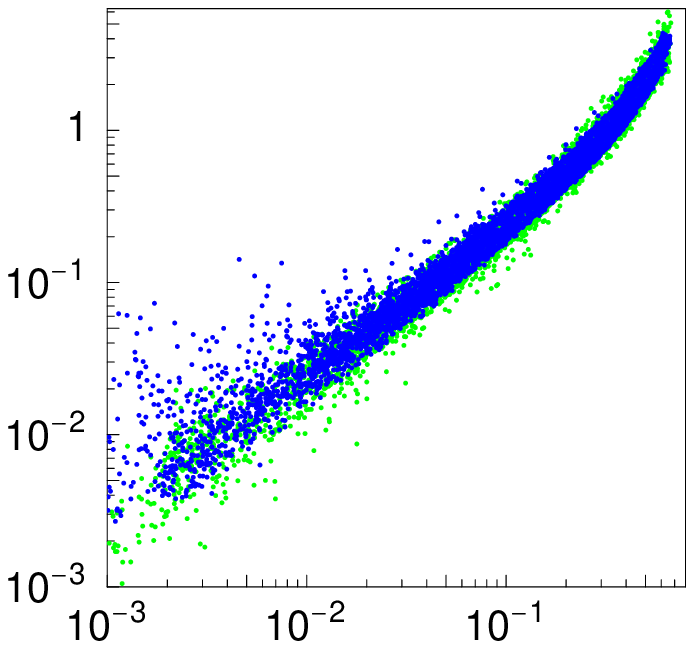,height=4.5cm,width=4.3cm}}}
\put(99,0){\mbox{\epsfig{
                   figure=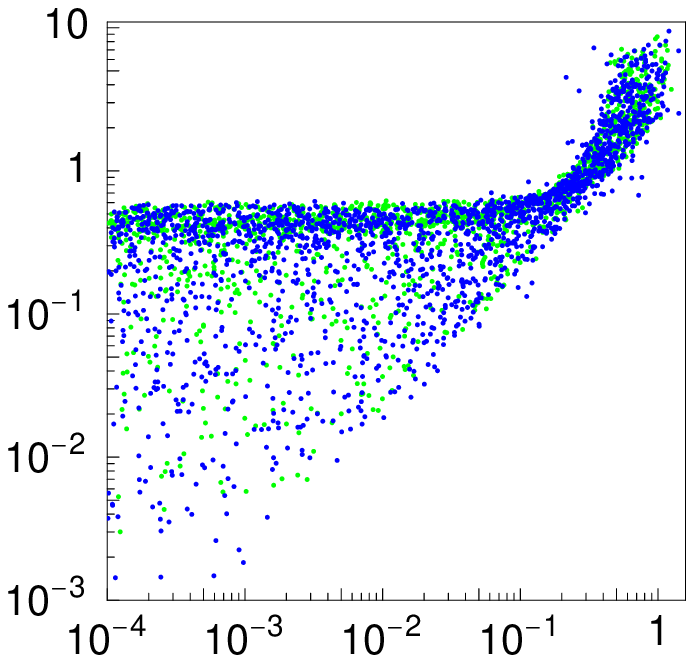,height=4.5cm,width=4.3cm}}}
\put(1,46){\mbox{\small BR($e q q'$)  /BR($\tau q q'$)}}
\put(30,-1){\mbox{$\tan^2 \theta_{atm}$}}
\put(52,46){\mbox{\small BR($e q q'$) /BR($\tau q q'$)}}
\put(88,-1){\mbox{$U^2_{e3}$}}
\put(102,46){\mbox{\small BR($e \tau \nu$) /BR($\mu \tau \nu$)}}
\put(130,-1){\mbox{$\tan^2 \theta_{sol}$}}
\end{picture}
\caption{Neutralino BR in bilinear \rp MSSM. \label{fig:BRchi0}}
\end{figure*}

There are a variety of implications of \rp
models~\cite{Allanach:1999bf}. The most obvious is that, unprotected
by any symmetry, the lightest supersymmetric particle (LSP), produced
with MSSM-like cross sections, will typically decay inside the
detector, as shown in \fig{fig:dchi0}, taken from~\cite{Bartl:2000yh}
Such decays are mainly into \texttt{visible modes}. Just as the
neutrino mixing angles characterizing the neutrino anomalies, in our
bilinear \rp MSSM model also the neutralino decay branching ratios are
determined by the same three fundamental \rp parameters in
\eq{eq:Wsuppot}. More exactly the neutrino mixing angles are
correlated with \texttt{ratios} of \rp parameters. These may be taken
as the $\Lambda_\mu/\Lambda_\tau$ for the atmospheric angle,
$\epsilon_e/\epsilon_\mu$ for the solar angle, and
$\Lambda_e/\Lambda_\tau$ for the angle which is probed by the reactor
experiments~\cite{chooz}. Here $\Lambda_i \equiv \epsilon_i \VEV{H_d}
+ \mu \VEV{\wt \nu_i}$, $\mu$ being the standard Higgsino mixing term.
As shown in ref.~~\cite{Hirsch:2000ef} due to the minimization
conditions the $\Lambda$ ratios do \texttt{not} introduce independent
parameters, hence the predictivity of the theory is manifest. As
\fig{fig:BRchi0} indicates, the LSP decay branching ratios are
strongly correlated with the leptonic mixing angles \footnote{The
  possibility of probing leptonic mixing angles at accelerator
  experiments in \rp models has been previously considered in refs.
  \cite{old2,Mukhopadhyaya:1998xj}.}.

\begin{figure*}
  \epsfxsize30pc
\includegraphics[width=0.45\textwidth,height=4.5cm]{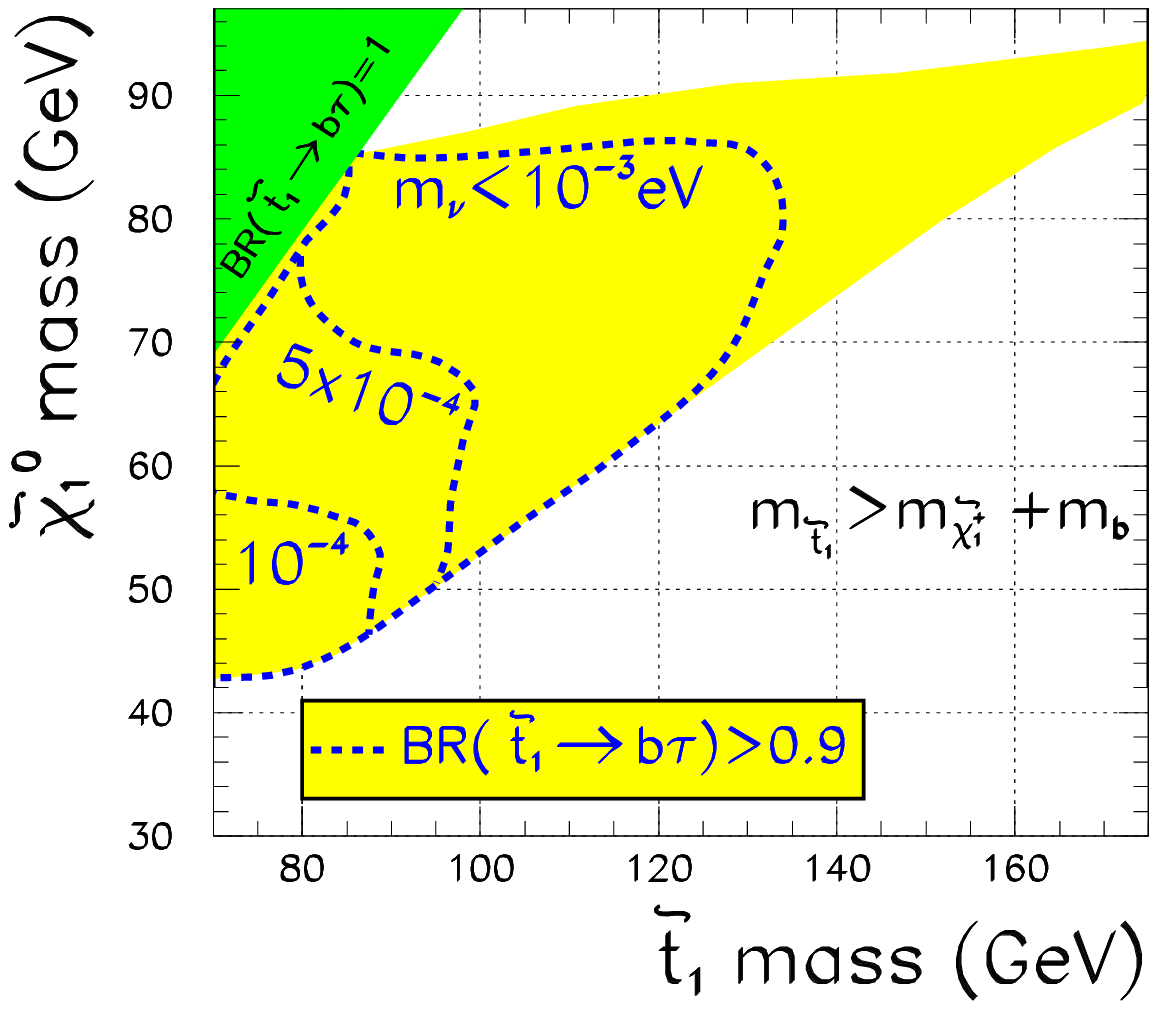} \hfil
\includegraphics[width=0.45\textwidth,height=4.5cm]{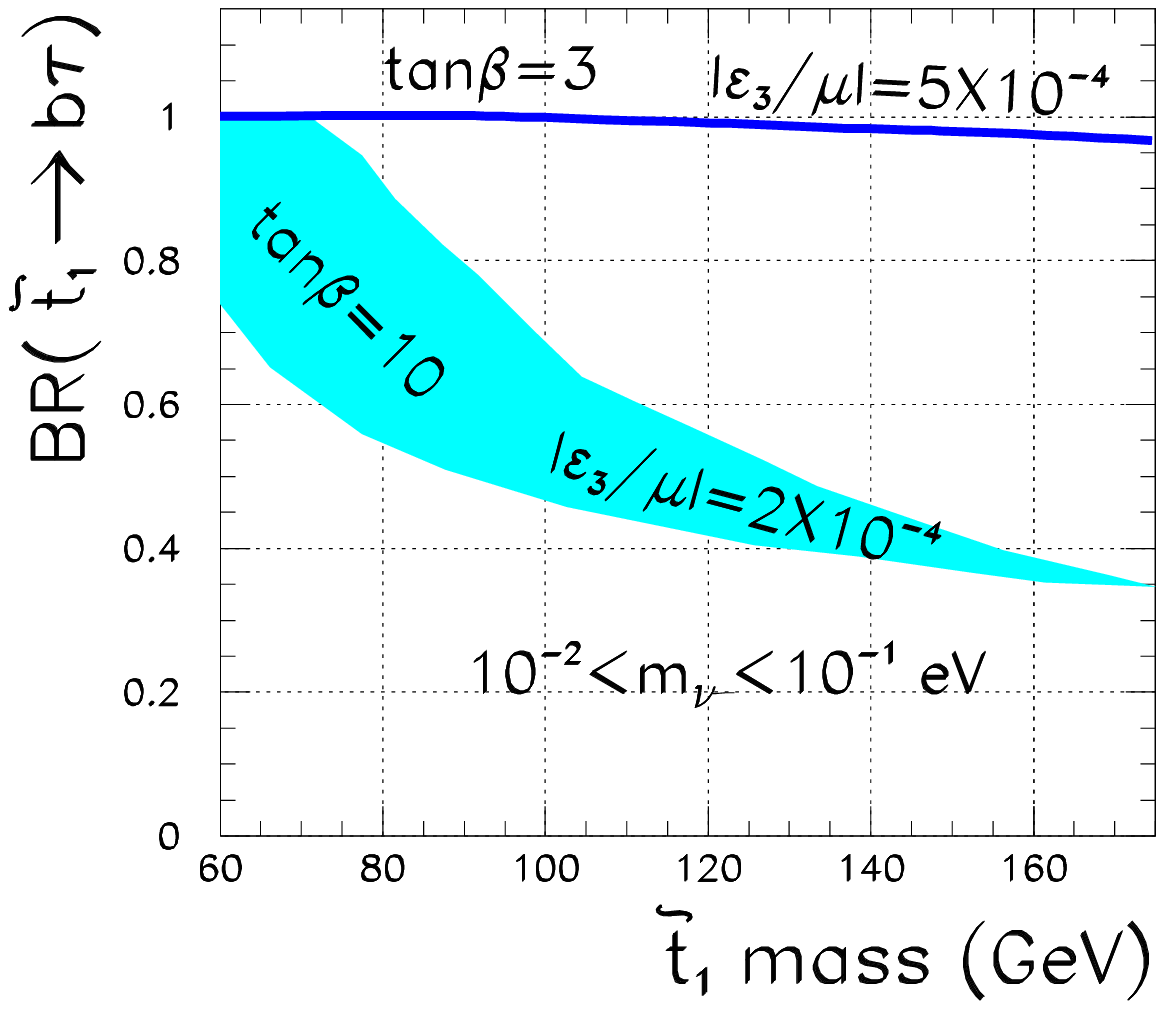}
\caption{Stop decay BR in bilinear \rp MSSM. \label{fig:BRstop}}
\end{figure*}

Neutralino decays can have remarkable consequences for gluino cascade
decays at the LHC, enhancing high lepton multiplicity event rates and,
correspondingly, thus decreasing the missing momentum signal expected
in the R-parity conserving MSSM~\cite{Bartl:1997cg}.
If R parity is broken particles other than the neutralino can be the
LSP.  One example is the stop~\cite{Bartl:1996gz}. In \fig{fig:BRstop}
we illustrate how two-body \rp decays of the lightest stop can be
sizeable when compared with standard decays~\cite{Diaz:2000ge}.
R parity violation can also affect gauge and Yukawa unification
\cite{Diaz:1999is}, texture predictions for $V_{cb}$
\cite{Diaz:2000wm} as well as $b \to s \gamma$ \cite{Diaz:1999wq}.
Turning to accelerators, \rp can affect the physics of the top
quark~\cite{Dreiner:1991dt} and it can lead to new signals for
chargino production at LEP2~\cite{deCampos:1999mf}, and affect the
phenomenology of supersymmetric scalars due to Higgs boson/slepton
mixing \cite{Akeroyd:1998iq}.

\noindent 
 
\section*{Acknowledgments}
This work was supported by DGICYT grant PB98-0693 and by the EEC under
the TMR contract ERBFMRX-CT96-0090. 
  M.H.~was supported by the
Marie-Curie program under grant No ERBFMBICT983000 and W.P.~by a
fellowship from the Spanish Ministry of Culture under the contract
SB97-BU0475382.

\end{document}